\documentclass[numreferences]{article}

\usepackage{amssymb}
\usepackage{amsfonts}
\usepackage{amsmath}
\usepackage{amstext}

\newenvironment{definitiontest}[1][Definition]{\noindent\textbf{#1.} }{\ \rule{0.5em}{0.5em}}

\begin{document}
\title{Algebras for Agent Norm-Regulation}

\author{Jan Odelstad \\ Division of Computer Science\\
           Univ of G\"avle, Sweden
E-mail: jod@hig.se
\and
Magnus Boman \\ Department of Computer and Systems Sciences\\
	The Royal Institute of Technology\\
	and\\
	Swedish Institute of Computer Science\\
        Kista, Sweden
E-mail: mab@sics.se}

\maketitle
\abstract{An abstract architecture for idealized multi-agent systems
whose behaviour is regulated by normative systems is developed and discussed.
Agent choices are determined partially by the preference ordering of possible states
and partially by normative considerations: The agent chooses that act
which leads to the best outcome of all permissible actions. If an action is
non-permissible depends on if the result of performing that action
leads to a state satisfying a condition which is forbidden, according to
the norms regulating the multi-agent system. This idea is formalized by
defining set-theoretic predicates characterizing multi-agent systems. 
The definition of the predicate
uses decision theory, the Kanger-Lindahl theory of normative positions, and
an algebraic representation of normative systems.}


\section{Introduction}

Within economic theory the consumer's behaviour has traditionally been
described as determined by a utility function. During the latest two decades
there has been a growing interest among researchers in how norms (for
example rules of law) give restrictions on the behaviour induced by the
utility function. The behaviour of the consumers or other economic agents,
according to this model, is the result of an interplay between optimization
of the utility function and restrictions due to norms. We will here show how
this model can be used for regulating the behaviour of artificial agents.

An important problem in the behavioral sciences is how the autonomy of the 
individual can be reconciled with collective
rationality, i.e. the rationality of the group?\footnote{%
Related to this problem is the question whether the autonomy of one
individual restricts the autonomy of other individuals.} If the individual
is a human agent and the collective in question is the state, we are
approaching a profound problem for politics and social science. However,
here we limit ourselves to some aspects of a more modest question: What is
an appropriate formal framework for the unification of individual autonomy
with collective rationality for \textit{artificial} agent systems? To our
technical problem, the solution could be an agent architecture with a large
range of applications. But this technical problem also has some bearing on
the more general one mentioned above, as it may make it possible to
construct and test model systems which resemble human social systems in
some basic sense.

The approach to unification of individual autonomy and collective
rationality chosen here is, as has already been emphasized, to focus on
norms. There are mainly two aspects of norms in relation to multi-agent systems
that will be dealt with, viz.

(1) the formal representation of norms regulating multi-agent systems and

(2) the role of norms in architectures for multi-agent systems.

These two aspects of multi-agent systems (\textsc{Mas}) are not independent, on the
contrary: they presuppose each other. The aim of the essay as far as norm
representation is concerned is to test an algebraic representation of
normative systems~\cite{LiOd99}~\cite{LiOd00}~\cite{LiOd02}~\cite{OdLi00}~\cite{OdLi02} 
utilizing the Kanger-Lindahl theory of normative 
positions~\cite{KaKa66}~\cite{Li77}~\cite{Li94}, and its later developments. In this
essay we will discuss some characteristic features of this kind of
norm representation and investigate the possibility of using it in \textsc{%
Mas} architectures. We have strived for imposing no particular choice
of \textsc{Mas} architecture; a related study for a Belief-Desire-Intention agent architecture,
for instance, is~\cite{DiMoSoCa00}.

Next a few words about (2). The role that norms will have in \textsc{Mas}
architectures is to delimit the autonomy of the agents. Metaphorically one
can say that the norms define the scope (\textit{Spielraum}) for an agent.
The \textquoteleft wish\textquoteright\ or \textquoteleft
desire\textquoteright\ of an agent is represented as a preference
structure over possible states or situations.\footnote{%
A preference structure consists of a preference relation and perhaps one or
several other relations, for example a difference relation.} From the norms
of the system will follow a deontic structure of the states (consisting 
for example of normative positions) implying that some states are
permissible while the rest are non-permissible. The agent chooses an act
which leads to the permissible state it prefers the most. In section 3, we
will transform this idea to a theory or model of how norms can be
used to regulate the behaviour of multi-agent systems. The model will use the
kind of norm representation mentioned above and it will be expressed in
terms of set-theoretical predicates. Since this is the first attempt to test
the above described approach we will use a rather idealized kind of \textsc{%
Mas}. The extent and goal of the idealization will be discussed at the
beginning of section 3.

The study of how the behaviour of agents can be regulated by norms has been
pursued within many fields, including biology, computer science, economics,
law, sociology and philosophy. Formalisms have often been employed in
cross-disciplinary fashion, as in the combination of mathematics and logic~%
\cite{OdLi00}, speech act theory and computational linguistics~\cite{Di99},
and computer science and anticipatory systems~\cite{BoDaYo99}. What
constitutes a norm naturally differs between such approaches. In this
essay, a norm is represented as an implicative sentence where the
antecedent is a descriptive condition stating the circumstances of an agent,
and the consequent is a condition expressing the normative position that the
agent has with respect to a state of affairs. A normative system is a
joining of two Boolean algebra based on such implications. Our approach to
norms is within the tradition of algebraic logic.

It is possible to regard the
expression \textquoteleft norms for artificial agents\textquoteright\
as metaphorical, but we do not. From a formal point of view,
multi-agent systems are sufficiently similar independently of whether the
agents are human or not. In this essay, we adopt a fairly formal notion of
a norm: a norm is characterized by its form or structure and not, for
example, by its function.

In section 2, norms and normative systems are examined from different
perspectives. The Kanger-Lindahl theory of normative positions is presented,
and the representation problem of normative systems is introduced and an
algebraic approach to its solution is outlined. In section 3, an abstract
architecture for \textsc{Mas} based on deontic-action-logic is
introduced. The formal structure of norms regulating the behaviour of such
systems is presented in section 4. In sections 2-4, some of the ideas are
elucidated using the waste collectors example. In section 5, an important
step in the theory construction is presented, namely a scheme for how
normative positions will restrict the set of acts that the agents are
permitted to choose from. Section 6 contains conclusions and a discussion of
possible extensions of the study.

\section{Representations of Norms}

\subsection{The Waste Collectors Example}

We illustrate some of the ideas in this essay using an example, introduced
in~\cite{Bo99}, of a team of robots collecting nuclear waste. Each robot is
here represented as one agent, and is placed on a spatial grid. It is
important to realize that this example is here exploited for elucidating the
role of norms in regulating the behavior of an agent system. It is not what
the norms say, the material content of the norms, that is of interest here
but their logical form and their formal connection to other parts of the
agent architecture. Thus it is not the reasonableness of the normative
systems which interest us but the reasonableness of the form of their
representation and their role in the architecture.

Think of a spatial grid of squares in rows and columns. Each square is
assigned a coordinate consisting of an ordered pair of integers, where
the first number in the pair represents the column and the second the row.
On some squares there is a number representing an amount of waste. A group
of agents, in the sequel called collectors, tries to collect as much waste as
possible. Each collector has a utility function, such that the utility
depends on the amount of waste per unit time (or unit distance).

The agents can move one square at a time in four directions, up (north),
down (south), left (west) and right (east); these are the possible actions
of the agents. But there are restrictions on how they may move. It is
especially important how an agent may move in the neighbourhood of another.
The \emph{protected sphere} around the agent $\omega $ consists of nine
squares forming a large square with $\omega $ in the middle, i.e. the
protected sphere is the von Neumann neighbourhood of $\omega $. The
protected spheres of $\omega _{1}$ and $\omega _{2}$ can overlap with 1, 2,
3, 4, 6 or 9 squares.

We will now present a set of norms which regulates
how the agents may move. Suppose $\omega _{1}$ is about to move.

\strut \noindent $(n_{1})$ If the protected spheres of $\omega _{1}$ and $%
\omega _{2}$ do not overlap then $\omega _{1}$ may not move so that $\omega
_{1}$ and $\omega _{2}$ overlap with two or three squares.

\strut \noindent $(n_{2})$ If the protected spheres of $\omega _{1}$ and $%
\omega _{2}$ do not overlap then $\omega _{1}$ may move so that $\omega _{1} 
$ and $\omega _{2}$ overlap with zero or one square.\footnote{%
Note that if the protected spheres of $\omega _{1}$ and $\omega _{2}$ do not
overlap then $\omega _{1}$ cannot move so that the spheres of $\omega _{1}$
and $\omega _{2}$ overlap with more than three squares.}

\strut \noindent $(n_{3})$ If the protected spheres of $\omega _{1}$ and $%
\omega _{2}$ overlap with one or two squares, then $\omega _{1}$ may move so
that the protected spheres of $\omega _{1}$ and $\omega _{2}$ overlap with
any number of squares (even zero).

\strut \noindent $(n_{4})$ $\omega _{1}$ may move so that the protected
spheres of $\omega _{1}$ and $\omega _{2}$ overlap with six squares only if
the protected spheres of $\omega _{1}$ and $\omega _{2}$ overlap with at
least four squares.

\strut \noindent $(n_{5})$ If the protected spheres of $\omega _{1}$ and $%
\omega _{2}$ overlap with four squares then $\omega _{1}$ shall move so
that the protected spheres of $\omega _{1}$ and $\omega _{2}$ do not overlap with
three squares.

\strut \noindent $(n_{6})$ If the protected spheres of $\omega _{1}$ and $%
\omega _{2}$ overlap with six squares, then $\omega _{1}$ must move so that
the protected spheres of $\omega _{1}$ and $\omega _{2}$ overlap with at
least four squares.

\strut \noindent $(n_{7})$ $\omega _{1}$ may never move so that the
protected spheres of $\omega _{1}$ and $\omega _{2}$ overlap with nine
squares.

\strut \noindent $(n_{8})$ $\omega _{1}$ may always move so that the
protected spheres of $\omega _{1}$ and $\omega _{2}$ overlap with zero
squares.

This is a very primitive normative system, but it is not so easy to see what
it says. To grasp the content of the normative system a transparent way of
structuring and representing the norms is necessary. It is easy to see that
most of the norms, but not all of them, are implications, i.e. of the form $%
p $ $implies$ $q$. However, as we will see in a later section, all of them
can easily be transformed to implications. Note also that the norms contain
expressions like \textit{may, may not} and \textit{shall}.
To simplify, let us introduce some predicates. The protected sphere around $%
\omega $ is denoted $Protec(\omega )$. $Lap_{i}(\omega _{1},\omega _{2})$
means that the protected spheres around $\omega _{1}$ and $\omega _{2}$
overlap with $i$ squares. As a first step in the formalization of norms $%
(n_{1})$ and $(n_{2})$ we can write them as follows:\smallskip

\noindent $(n_{1})$ If $Lap_{0}(\omega _{1},\omega _{2})$ then it may not be
the case that $\omega _{1}$ sees to it that $Lap_{2}(\omega _{1},\omega
_{2}) $ or $Lap_{3}(\omega _{1},\omega _{2})$.\smallskip

\noindent $(n_{2})$ If $Lap_{0}(\omega _{1},\omega _{2})$ then it may be the
case that $\omega _{1}$ sees to it that $Lap_{0}(\omega _{1},\omega _{2})$
and $Lap_{1}(\omega _{1},\omega _{2})$.\smallskip

The expression `$\omega _{1}$ sees to it that'\ states the fact that $\omega
_{1}$ is active and moves and thereby sees to it that the overlap in
question is realized. Let us use Do($\omega _{1},F$) as an abbreviation for `%
$\omega _{1}$ sees to it that $F$'. Furthermore, let us use `May' as an
abbreviation for `it may be the case that' and `Shall' for `it shall be the case that.
We can then represent norms $(n_{1})$ and 
$(n_{2})$ as follows.\smallskip

\noindent $(n_{1})$ $Lap_{0}(\omega _{1},\omega _{2})$ $\longrightarrow $ $%
\lnot $May Do$(\omega _{1},Lap_{2}(\omega _{1},\omega _{2})\vee
Lap_{3}(\omega _{1},\omega _{2}))$\smallskip

\noindent $(n_{2})$ $Lap_{0}(\omega _{1},\omega _{2})$ $\longrightarrow $
May Do$(\omega _{1},Lap_{0}(\omega _{1},\omega _{2}))\wedge $May Do$(\omega
_{1},Lap_{1}(\omega _{1},\omega _{2}))$\smallskip\ \newline
The other norms can be represented in a similar way.

\noindent $(n_{3})$ For all $i,0\leq i\leq 9,$ $Lap_{1}(\omega _{1},\omega
_{2})\vee Lap_{2}(\omega _{1},\omega _{2})\longrightarrow $ May Do$(\omega
_{1},Lap_{i}(\omega _{1},\omega _{2}))$\smallskip

\noindent $(n_{4})$ $\lnot Lap_{4}(\omega _{1},\omega _{2})\wedge \lnot
Lap_{6}(\omega _{1},\omega _{2})\wedge \lnot Lap_{9}(\omega _{1},\omega
_{2}) $ $\longrightarrow $ $\lnot $May Do$(\omega _{1},Lap_{6}(\omega
_{1},\omega _{2}))$\smallskip

\noindent $(n_{5})$ $Lap_{4}(\omega _{1},\omega _{2})$ $\longrightarrow $
Shall Do$(\omega _{1},\lnot $ $Lap_{3}(\omega _{1},\omega _{2}))$\smallskip

\noindent $(n_{6})$ $Lap_{6}(\omega _{1},\omega _{2})$ $\longrightarrow $
Shall Do$(\omega _{1},Lap_{4}(\omega _{1},\omega _{2})\vee Lap_{6}(\omega
_{1},\omega _{2})\vee Lap_{9}(\omega _{1},\omega _{2}))$\smallskip

\noindent $(n_{7})$ $\omega _{1}\neq \omega _{2}\longrightarrow \lnot $May Do%
$(\omega _{1},Lap_{9}(\omega _{1},\omega _{2}))$\smallskip

\noindent $(n_{8})$ May Do$(\omega _{1},Lap_{0}(\omega _{1},\omega _{2}))$%
\smallskip

In the formulations of $(n_{1})-(n_{8})$ we have as a simplification
omitted the universal quantifiers. For example $(n_{1})$ ought to be
written more completely as\smallskip

\noindent $(n_{1}^{\prime })$ $\forall \omega_{2}:
[Lap_{0}(\omega _{1},\omega _{2})$ $\longrightarrow $ $\lnot $May Do$%
(\omega _{1},Lap_{2}(\omega _{1},\omega _{2})\vee Lap_{3}(\omega _{1},\omega
_{2}))]$\smallskip

This is just the first step in the representation of norms using deontic and
action logic. In the next section we will discuss this kind of
representation in some detail and show how we can get a more effective
representation. We will return to the waste collectors in sections 3 and 4.

Note that a typical norm asserts that if a condition on the position of $%
\omega _{1}$ is fulfilled, then $\omega _{1}$ may or may not see to it that
another condition will or will not be the case. An idea elaborated on in
section 5 is that it is not permissible for $\omega _{1}$ to perform an act
which leads to a situation such that a condition $c$ will be fulfilled, if $c
$ is such that $\omega _{1}$ may not see to it that $c$ is fulfilled. In
this way the norms formulated in terms of requirements on positions
determine what is permissible or obligatory for the agent to do in a given
situation.

\subsection{Norms as Ordered Pairs}

A typical norm is a universal implication and in predicate logic it can
often be represented as a universal sentence of the following form:

\noindent $(n_{9}):\forall x_{1},...,x_{\nu }:p(x_{1},...,x_{\nu
})\longrightarrow q(x_{1},...,x_{\nu }).\smallskip $

Syntactically it consists of three parts: the sequence of universal
quantifiers, the antecedent formula and the consequent formula. Note that
norm ($n_{9}$) correlates open sentences: $p(x_{1},...,x_{\nu })$ is
correlated to $q(x_{1},...,x_{\nu })$. An alternative point of view is to
regard $p$ and $q$ as \emph{conditions} and a norm as a relational statement
correlating them:

\noindent $(n_{10}):p\mathcal{R}q\smallskip $

It is important here that the free variables in $p(x_{1},...,x_{\nu })$ are
the same and in the same order as the free variables in $q(x_{1},...,x_{\nu
}).$ (It is, however, not necessary that $p$ and $q$ have the same arity,
for details see~\cite{LiOd00}.) $\mathcal{R}$ is a binary
relation, and $p\mathcal{R}q$ is a relational statement equivalent to $%
\left\langle p,q\right\rangle \in \mathcal{R}.$ Thus, we can represent a
norm as $p\mathcal{R}q$ or $\left\langle p,q\right\rangle \in \mathcal{R}$,
and from the latter relational statement it is just a small step to the
representation of $(n_{10})$ as the ordered pair $\left\langle
p,q\right\rangle $ where $p$ is called the \emph{ground} and $q$ the \emph{%
consequence} of the norm. A ground is a descriptive and a consequence is a
normative condition.

As is easy to see, we can form conjunctions, disjunctions and negations of
conditions in the following way.$\smallskip $

$(p\wedge q)(\omega _{1},...,\omega _{\nu })$ if and only if $p(\omega
_{1},...,\omega _{\nu })$ and $q(\omega _{1},...,\omega _{\nu }).$

$(p\vee q)(\omega _{1},...,\omega _{\nu })$ if and only if $p(\omega
_{1},...,\omega _{\nu })$ or $q(\omega _{1},...,\omega _{\nu }).$

$(p^{\prime })(\omega _{1},...,\omega _{\nu })$ if and only if $\lnot
p(\omega _{1},...,\omega _{\nu }).\smallskip $

It is therefore possible to construct Boolean algebras of conditions and we
will return to this in subsection 2.4.

If $p$ is a $\nu $-ary condition and $i_{1}$,...,$i_{\nu }$ are individuals,
then $p(i_{1},...,i_{\nu })$ is a sentence. A norm $\left\langle
p,q\right\rangle $ can be regarded as a mechanism of inference. We can
distinguish two cases. Suppose that $p$ and $s$ are descriptive conditions
and $q$ and $t$ are normative.

\begin{enumerate}
\item From $p(i_{1},...,i_{\nu })$ together with $\left\langle
p,q\right\rangle $ follows $q(i_{1},...,i_{\nu }).$
\item From $s\mathcal{R}p$ together with $\left\langle p,q\right\rangle $
and $q\mathcal{R}t$ follows $s\mathcal{R}t.$\footnote{%
Note that $s\mathcal{R}p$ is relating two sentences of the same kind and the
same holds for $q\mathcal{R}t$; $s$ and $p$ are descriptive but $q$ and $t$
are normative. A norm consists of sentences of different kinds.}
\end{enumerate}

In 1, $\left\langle p,q\right\rangle $ functions as a deductive mechanism
correlating sentences by means of instantiation, while in 2, $\left\langle
p,q\right\rangle $ correlates one condition to another condition. Therefore,
in the terminology of~\cite{AlBu71}, 
1 corresponds to the correlation of individual cases to individual
solutions, and 2 corresponds to the correlation of generic
cases to generic solutions.

Let us return to the suggested norm $(n_{1})$ in the waste
collector example above. The ground of the norm is the binary condition $%
Lap_{0}$ and the consequence is the condition $\Lambda$ defined in the following
way:$\smallskip $

$\Lambda(\omega _{1},\omega _{2})$ iff $\lnot $May Do$(\omega _{1},Lap_{2}(\omega _{1},\omega
_{2})\vee Lap_{3}(\omega _{1},\omega _{2})).\smallskip $

$\Lambda$ is a normative condition obtained by applying the operators
May and Do to the disjunction of the descriptive conditions $Lap_{2}$ and $%
Lap_{3}.$ 

\subsection{Normative Positions}

An important contribution to deontic logic was made by Stig Kanger, who
combined the deontic operator Shall with the binary action operator Do~\cite%
{Ka57}. To be specific, Shall Do($x,q$) means that it shall be that $x$ sees
to it that $q$, while $\lnot $Shall Do($y,\lnot q$) means that it is not the
case that it shall be that $y$ sees to it that not $q$. The combination of
the deontic operator Shall with the action operator Do and the negation
operation $\lnot $ gives us a powerful language for expressing purely
normative sentences. Kanger emphasized the possibilities of external and
internal negation of sentences where these operators are combined. Using
combinations of deontic and action operators, we can formulate norms in a
concise way. A conditional norm may for example have the form:\smallskip

$p(x,y)\rightarrow $ Shall Do($x,\lnot q(y)$).\smallskip

Note that the sentence May Do($x,q$) can be defined in terms of the
operators Shall and Do in the following way:\smallskip

May Do($x,q$) if and only if $\lnot $Shall $\lnot $Do($x,q$).\smallskip

The logical postulates for Shall and Do assumed by Kanger are (where $%
\implies $ is the relation of logical consequence and $\iff $ of logical
equivalence):

\begin{enumerate}
\item If $p\implies q$, then Shall $p\implies $ Shall $q$.

\item (Shall $p$ \& Shall $q$) $\implies $ Shall($p$ \& $q$).

\item Shall $p\implies \lnot $Shall $\lnot $\ $p$.

\item If $p\iff q$, then Do($x,p$)$\iff $Do($x,q$).

\item Do($x,p$) $\implies p$.
\end{enumerate}

The five conditions are, according to Kanger, acceptable in most contexts.

Kanger used the deontic-action-language as a basis for a theory of normative
positions and his theory, expressed as a theory of types of rights, was
further developed by Lars Lindahl in his three systems of types of normative
positions~\cite{Li77}. The simplest one is the system of one-agent types of
normative position, and we will restrict ourselves to the utilization of this
system here.

Let $\pm \alpha $ stand for either of $\alpha $ or $\lnot \alpha .$ Starting
from the scheme $\pm $May$\pm $Do($x,\pm q),$ where $\pm $ stands for the
two alternatives of affirmation or negation, a list is made of all maximal
and consistent conjunctions---maxiconjunctions (see ~\cite{Ma86} p.405f.)---
such that each conjunct satisfies the scheme. Note that the expression $%
\lnot $Do$(x{,}q) $\& $\lnot $Do($x,\lnot q)$ expresses \textit{x}'s
passivity with regard to $q$. Here this expression is abbreviated as Pass($%
x,q$). By this procedure the following list of seven maxiconjunctions is
obtained, which are denoted \textbf{T}$_{1}$($x,q$),\dots ,\textbf{T}$_{7}$($%
x,q$) (see ~\cite{Li77}, p.92).

\begin{itemize}
\item \textbf{T}$_{1}(x,q):\;$MayDo$(x,q)\;\&\;$MayPass$(x,q)\;\&\;$MayDo$%
(x,\lnot q).$

\item \textbf{T}$_{2}(x,q):\;$MayDo$(x,q)\;\&\;$MayPass$(x,q)\;\&\;\lnot $%
MayDo$(x,\lnot q).$

\item \textbf{T}$_{3}(x,q):\;$MayDo$(x,q)\;\&\;\lnot $MayPass$(x,q)\;\&\;$%
MayDo$(x,\lnot q).$

\item \textbf{T}$_{4}(x,q):\;\lnot $MayDo$(x,q)\;\&\;$MayPass$(x,q)\;\&\;$%
MayDo$(x,\lnot q).$

\item \textbf{T}$_{5}(x,q):\;$MayDo$(x,q)\;\&\;\lnot $MayPass$%
(x,q)\;\&\;\lnot $MayDo$(x,\lnot q).$

\item \textbf{T}$_{6}(x,q):\;\lnot $MayDo$(x,q)\;\&\;$MayPass$%
(x,q)\;\&\;\lnot $MayDo$(x,\lnot q).$

\item \textbf{T}$_{7}(x,q):\;\lnot $MayDo$(x,q)\;\&\;\lnot $MayPass$%
(x,q)\;\&\;$MayDo$(x,\lnot q).$
\end{itemize}

\textbf{T}$_{1}$,\dots ,\textbf{T}$_{7}$ are called the types of one-agent
positions. Note that $\lnot $MayDo $(x,q)\;\&\;\lnot $ MayPass $%
(x,q)\;\&\;\lnot $MayDo$(x,\lnot q)$ is logically false, according to the
logic of Shall and May. It is easy to see that the last three types can more
concisely be described as follows:

\begin{itemize}
\item \textbf{T}$_{5}(x,q):\;$Shall Do$(x,q).$

\item \textbf{T}$_{6}(x,q):\;$Shall Pass$(x,q).$

\item \textbf{T}$_{7}(x,q):\;$Shall Do$(x,\lnot q).$
\end{itemize}

Note that the following symmetry principles hold:

\begin{itemize}
\item \textbf{T}$_{1}(x,q)$ if and only if \textbf{T}$_{1}(x,\lnot q)$

\item \textbf{T}$_{2}(x,q)$ if and only if \textbf{T}$_{4}(x,\lnot q)$

\item \textbf{T}$_{3}(x,q)$ if and only if \textbf{T}$_{3}(x,\lnot q)$

\item \textbf{T}$_{5}(x,q)$ if and only if \textbf{T}$_{7}(x,\lnot q)$

\item \textbf{T}$_{6}(x,q)$ if and only if \textbf{T}$_{6}(x,\lnot q)$
\end{itemize}

The systems of normative positions can serve as a tool for describing the
normative positions of different agents $x,y,z...$ with regard to states of
affairs $p,q,r,....$ A set of such descriptions, however, is not a
representation of a normative system. This is due to the fact that a
normative system is not a description of the actual normative positions of
individuals. Rather, the essential feature of a normative system consists in
so-called normative correlations, i.e., as jurists might say, in correlating
normative consequences to operative facts. The formal system of normative
positions increases the expressive power of norm formulation as regards
consequences.

The one-agent types in the Kanger-Lindahl theory of normative positions can
be used as operators on descriptive conditions to get deontic conditions~%
\cite{LiOd02}. As a simple example, suppose that $r$ is a unary condition.
Then $T_{i}r$ (with $1\leq i\leq 7)$ is the binary condition such that $%
T_{i}r(y,x)\text{ iff \textbf{T}}_{i}(x,r(y))$, where \textbf{T}$%
_{i}(x,r(y)) $ is the $i$th formula of one-agent normative positions. Note
that for example \textbf{T}$_{3}(x,r(y))$ means

\begin{center}
MayDo$(x,r(y))\;\&\;\lnot $MayPass$(x,r(y))\;\&\;$MayDo$(x,\lnot r(y)).$
\end{center}

If $\left\langle p,T_{i}r\right\rangle $ is a norm, then from $%
p(x_{1},x_{2}) $ we can, by using the norm, infer $T_{i}r(x_{1},x_{2})$ and
thus also \textbf{T}$_{i}(x_{2},r(x_{1}))$, which means that, with regard to
the state of affairs $r(x_{1})$, $x_{2}$ has a normative position of type 
\textbf{T}$_{i}.$

The theory of normative positions was developed during the 60s and 70s,
primarily as an analytical tool to be used in jurisprudence and political
science; Kanger's theory was originally expressed as a theory of types of
rights. Some further refinement of the systems have more recently been made
by Andrew J.I. Jones and Marek Sergot (see \cite{JoSe93}~\cite{JoSe96}~\cite{Se99}~\cite{Se01}). 
A special feature of the work of
Jones and Sergot, as of Herrestad and Krogh~\cite{Kr95}~\cite{KrHe99}, is that applications in
computer science are in view. Even though Sergot has constructed a
computer program, Norman-G, based on the theory of normative positions,
we have chosen to base our exposition on Kanger/Lindahl in order to not
complicate the model further.

\subsection{Representations of Norms: The Algebraic Approach}

An adequate representation of norms is important for the construction of
norm-based architectures and we will here use an algebraic approach to the
representation problem.\footnote{%
This subsection is based on earlier work by Lindahl and Odelstad, 
see~\cite{LiOd02}~\cite{OdLi00}~\cite{OdLi02}.} 
One of the main tools in this endeavour is the theory of a
Boolean quasi-ordering, which is an extension of the theory of Boolean
algebras. A norm is regarded as consisting of two objects, a ground and a
consequence standing in a relation to each other. The ground belongs to one
Boolean quasi-ordering and the consequence to another. Therefore, we can
view a normative system as a joining of a Boolean quasi-ordering of grounds
to a Boolean quasi-ordering of consequences.\smallskip

\begin{definitiontest}
The relational structure $\langle B,\wedge ,^{\prime} ,R\rangle $ is a \emph{%
Boolean quasi-ordering (Bqo) }if $\langle B,\wedge ,^{\prime} \rangle $ is
Boolean algebra and $R$ is a binary, reflexive and transitive relation on $B$
(i.e. $R$ is a quasi-ordering) such that $R$ satisfies the following
conditions for all $a$, $b$ and $c$ in $B$:\newline
(1) $aRb$ and $aRc$ implies $aR(b\wedge c)$.\newline
(2) $aRb$ implies $b^{\prime }Ra^{\prime }$.\newline
(3) $(a\wedge b)Ra$.\newline
(4) not $\mathbf{\top }R\bot .$
\end{definitiontest}

Before we describe the use of Boolean quasi-orderings for representing
norms, let us say a few words on some formal aspects of such structures. The
indifference part of $R$ is denoted $Q$ and is defined by: $aQb$ if and only
if $aRb$ and $bRa$. Similarly, the strict part of $R$ is denoted $S$ and is
defined by: $aSb$ if and only if $aRb$ and not $bRa$.

Let $\leq $ be the partial ordering determined by the Boolean algebra $%
\langle B,\wedge ,^{\prime} \rangle .\footnote{%
As per usual, $\leq $ is defined by $a\leq b$ if and only if $a\wedge b=a$.}$
From requirement (3) for Boolean quasi-orderings it follows that $a\leq b$
implies $aRb$. If $\left\langle B,\wedge ,^{\prime },R\right\rangle $ is a
Boolean quasi-ordering then we say that the Boolean algebra $\left\langle
B,\wedge ,^{\prime }\right\rangle $ is the \emph{reduct} of $\left\langle
B,\wedge ,^{\prime },R\right\rangle $, denoted $\mathcal{B}^{red}.$

Suppose that $\mathcal{B}=\langle B,\wedge ,^{\prime} ,R\rangle $ is a Boolean
quasi-ordering and $Q$ is the indifference part of $R$. The \emph{quotient
algebra} of $\mathcal{B}$ with respect to $Q$ is a structure $\langle
B/Q,\cap ,-,\preceq _{Q}\rangle $ such that $\langle B/Q,\cap ,-\rangle $ is
a Boolean algebra and $\preceq _{Q}$ is the partial ordering determined by
this algebra. The natural mapping of $\langle B,\wedge ,^{\prime} \rangle $
onto $\langle B/Q,\cap ,-\rangle $ is a homomorphism~\cite{OdLi00}, and $%
\langle B/Q,\cap ,-\rangle $ is called the \emph{quotient reduction }of $%
\mathcal{B}.$ Thus there are two Boolean algebras which should be kept
apart, namely $\mathcal{B}^{red},$ i.e. the reduct of $\mathcal{B},$ and the
quotient reduction of $\mathcal{B}.$

Although, by a transition to equivalence classes, from a Boolean
quasi-ordering we get a new Boolean algebra, there is a point in remaining
within the framework of Boolean quasi-orderings. In the models where the
domain of a Boolean quasi-ordering is a set of conditions, we may want to
distinguish two conditions $a$ and $b$ even though it holds that $aQb$ (and
therefore $a$ and $b$ belong to the same $Q$-equivalence class), because
they may have different meaning.

An important class of models of the theory of Boolean quasi-orderings
consists of models having a set of conditions as its domain.\smallskip

\begin{definitiontest}
A \emph{condition implication structure (cis)} is a Boolean quasi-ordering%
\emph{\ }$\mathcal{B}=\langle B,\wedge ,^{\prime },R\rangle $ such that $B$
is a domain of conditions, and $R$ is such that $aRb$ represents that 
$a$ \emph{implies} $b$.
\end{definitiontest}\bigskip

This reading of $R$ is justified since, if $a$ and $b$ are $\nu -$ary
conditions, $aRb$ is the representation of 
\begin{equation*}
\forall x_{1},..,x_{\nu }:a(x_{1},...,x_{\nu })\rightarrow
b(x_{1},...,x_{\nu }).
\end{equation*}

The theory of Boolean quasi-orderings is of a very general character, and
condition implication structures are not the only kind of models which are
interesting as representations of normative structures. It is easy to see
that we can construct a model of this theory out of a first order theory $%
\Sigma $. Consider the structure $\langle B,\wedge ,^{\prime },R\rangle $
where $\langle B,\wedge ,^{\prime }\rangle $ is the Lindenbaum algebra of
the predicate calculus. Let $R$ be the quasi-ordering on $B$ determined by
the Lindenbaum algebra of $\Sigma .$ Then $\langle B,\wedge ,^{\prime
},R\rangle $ is a Boolean quasi-ordering (cf.~\cite{BeSl69}, p.61 and~\cite{ChKe73}, p.73).

From an algebraic point of view a norm is a kind of link or joining of one
Boolean quasi-ordering to another. To make this idea precise we need some
definitions.\smallskip

\begin{definitiontest}
The \emph{subinterval} relation generated by the quasi-orderings $%
\left\langle B_{1},R_{1}\right\rangle $ and $\left\langle
B_{2},R_{2}\right\rangle $ is the binary relation $\trianglelefteq $ on $%
B_{1}\times B_{2}$ such that $\langle a_{1},a_{2}\rangle \trianglelefteq
\langle b_{1},b_{2}\rangle $ if and only if $b_{1}Ra_{1}$ and $a_{2}Rb_{2}.$ 
\end{definitiontest}\bigskip

Note that $\trianglelefteq $\ is a quasi-ordering, i.e. transitive and
reflexive. Let $\Bumpeq $ denote the equality part of $\trianglelefteq $ and 
$\vartriangleleft $ the strict part of $\trianglelefteq $. Then the
following holds:\medskip

$\langle a_{1},a_{2}\rangle \Bumpeq \langle b_{1},b_{2}\rangle $ if and only
if $b_{1}Q_{1}a_{1}$ and $a_{2}Q_{2}b_{2}$

$\langle a_{1},a_{2}\rangle \vartriangleleft \langle b_{1},b_{2}\rangle $ if
and only if $(b_{1}S_{1}a_{1}$ and $a_{2}R_{2}b_{2})$ or $(b_{1}R_{1}a_{1}$
and $a_{2}S_{2}b_{2})$\medskip

\noindent where $Q_{i}$ is the equality-part of $R_{i}$ and $S_{i}$ is the
strict part of $R_{i}$.
$\langle a_{1},a_{2}\rangle $ is a \emph{minimal element} in $X\subseteq
B_{1}\times B_{2}$ with respect to $\left\langle B_{1},R_{1}\right\rangle $
and $\left\langle B_{2},R_{2}\right\rangle $ if and only if there is no $%
\langle x_{1},x_{2}\rangle \in X$ such that $\langle x_{1},x_{2}\rangle
\vartriangleleft \langle a_{1},a_{2}\rangle .$

A quasi-ordering is closely related to a partial ordering. If $\langle
B,R\rangle $ is a quasi-ordering and $Q$ is the equivalence part of $R$,
then $R$ generates a partial ordering on the set of $Q$-equivalence classes
generated from $B$. The definitions of least upper bound (lub) and greatest
lower bound (glb) for partial orderings are easily extended to
quasi-orderings, but the lub or glb of a subset of a quasi-ordering is not
necessarily unique but can consist of a set of elements.\smallskip

\begin{definitiontest}
A \emph{joining-system} is an ordered triple $\left\langle \mathcal{B}_{1},%
\mathcal{B}_{2},\mathcal{J}\right\rangle $ such that $\mathcal{B}%
_{1}=\langle B_{1},R_{1}\rangle $ and $\mathcal{B}_{2}=\langle
B_{2},R_{2}\rangle $ are quasi-orderings and $\mathcal{J}\subseteq
B_{1}\times B_{2}$ and the following conditions are satisfied where $%
\trianglelefteq $\ is the subinterval relation generated by $\mathcal{B}_{1}$
and $\mathcal{B}_{2}$:\newline
(1) for all $b_{1},c_{1}\in B_{1}$ and $b_{2},c_{2}\in B_{2},$ $%
\left\langle b_{1},b_{2}\right\rangle \in \mathcal{J}$ and $\left\langle
b_{1},b_{2}\right\rangle \trianglelefteq \left\langle
c_{1},c_{2}\right\rangle $ implies $\left\langle c_{1},c_{2}\right\rangle
\in \mathcal{J},$\newline
(2) if for all $c_{1}\in C_{1}\subseteq B_{1},$ $\left\langle
c_{1},b_{2}\right\rangle \in \mathcal{J}$ and \emph{lub} $C_{1}\neq
\varnothing ,$ then $\left\langle a_{1},b_{2}\right\rangle \in \mathcal{J}$
for all $a_{1}\in $ \emph{lub} $C_{1},$\newline
(3) if for all $c_{2}\in C_{2}\subseteq B_{2},$ $\left\langle
b_{1},c_{2}\right\rangle \in \mathcal{J}$ and \emph{glb} $C_{2}\neq
\varnothing ,$ then $\left\langle b_{1},a_{2}\right\rangle \in \mathcal{J}$
for all $a_{2}\in $ \emph{glb} $C_{2}.$
\end{definitiontest}\bigskip

\begin{definitiontest}
Suppose that $\left\langle \mathcal{B}_{1},\mathcal{B}_{2},\mathcal{J}%
\right\rangle $ is a joining system. A \emph{minimal element}\ in $%
\left\langle \mathcal{B}_{1},\mathcal{B}_{2},\mathcal{J}\right\rangle $ is a
minimal element $\langle a_{1},a_{2}\rangle $\ in $\mathcal{J}$ with respect
to $\mathcal{B}_{1}$\ and $\mathcal{B}_{2}.$ The set of minimal elements in $%
\left\langle \mathcal{B}_{1},\mathcal{B}_{2},\mathcal{J}\right\rangle $ is
denoted $min\left\langle \mathcal{B}_{1},\mathcal{B}_{2},\mathcal{J}%
\right\rangle $ or just $min\mathcal{J}$.
\end{definitiontest}\bigskip

\begin{definitiontest}
A \emph{Boolean joining-system }$\left\langle \mathcal{B}%
_{1},\mathcal{B}_{2},\mathcal{J}\right\rangle $ is a joining-system such that $\mathcal{B}_{1}$
and $\mathcal{B}_{2}$ are Boolean quasi-orderings. A \emph{%
ground-consequence system} (\emph{gc-system}) is a Boolean joining-system $%
\left\langle \mathcal{B}_{1},\mathcal{B}_{2},\mathcal{J}\right\rangle $ such
that $\mathcal{B}_{1}$ and $\mathcal{B}_{2}$ are condition implication
structures.
\end{definitiontest}\bigskip

\begin{definitiontest}
A Boolean joining-system $\left\langle \mathcal{B}_{1},\mathcal{B}_{2},\mathcal{J}%
\right\rangle $ satisfies \emph{connectivity} if whenever $\left\langle
c_{1},c_{2}\right\rangle \in \mathcal{J}$ there is $\langle
b_{1},b_{2}\rangle \in \mathcal{J}$ such that $\langle b_{1},b_{2}\rangle $
is a minimal element in $\left\langle \mathcal{B}_{1},\mathcal{B}_{2},%
\mathcal{J}\right\rangle $ and $\left\langle b_{1},b_{2}\right\rangle
\trianglelefteq \left\langle c_{1},c_{2}\right\rangle .$
\end{definitiontest}\bigskip

It is easy to see that if a Boolean joining-system satisfies connectivity,
then the set of minimal joinings determines the set of joinings. If in a
Boolean joining-system $\left\langle \mathcal{B}_{1},\mathcal{B}_{2},%
\mathcal{J}\right\rangle $, $\mathcal{B}_{1}$ and $\mathcal{B}_{2}$ are
complete in a sense which is a straightforward generalization of the notion
of completeness applied to Boolean algebras, then $\left\langle \mathcal{B}%
_{1},\mathcal{B}_{2},\mathcal{J}\right\rangle $ satisfies connectivity. For
a proof, see \cite{Od03}.

A normative system $\mathcal{N}$ can be represented as a \emph{gc-}system $%
\left\langle \mathcal{B}_{1},\mathcal{B}_{2},\mathcal{J}\right\rangle.$\
The elements in $\mathcal{J}$ are then the norms of the system and in a norm 
$\left\langle a_{1},a_{2}\right\rangle,$ $a_{1}$ is the ground and $a_{2}$
the consequence. The elements in $min\mathcal{J}$\ constitute the set of minimal
norms of $\mathcal{N}$. If $\mathcal{N}$ satisfies connectivity, then it is
characterized by its set of minimal norms, a fact which will be of interest
in the sequel.

The set of minimal elements of a joining system exhibits an interesting
structure and it is possible to distinguish between different types of
joining systems depending on the properties of the set of minimal elements
(see~\cite{OdLi00} and~\cite{LiOd02}). It is
therefore also possible to distinguish between different kinds of
normative system depending on the structural properties of the set of
minimal norms. Furthermore, applications of the set of minimal norms seem
to make changes and extensions of normative systems easy to describe and
make it possible to divide the normative system into different parts
which can be changed independently. However, these lines of thought will not
be pursued here.

\section{An Architecture for Norm-Regulation of Multi-Agent Systems}

\subsection{Deontic-Action-Logic Multi-Agent Systems}

In this section we will give a definition of a deontic-action-logic based
multi-agent system, abbreviated \textsc{Dalmas}. \textsc{Dalmas} is an abstract
architecture (cf.~\cite{Wo02}, p.31.) for
idealized multi-agent systems using normative systems. The idea behind the
architecture is roughly the following. When it is agent $\omega $'s turn to
move it chooses an act out of a set of feasible alternatives and the result
will be that the system enters a new state; which state depends on the
actual state of the system when the act is performed. The agent's choice is
determined partially by the preference ordering of the possible states and
partially by the deontic structure: the agent chooses that act which leads
to the best outcome of all permissible actions. If an action is permissible
or not depends on whether the result of performing the action leads to a
state which satisfies a condition which is forbidden according to the
normative system regulating the multi-agent system. In this section we shall
see how this idea can be formalized.

\textsc{Dalmas} is a global clock (synchronous update), global state,
global dynamics system. It can be viewed as a simplification constructed for
conceptual and computer simulation purposes. In particular, we use the
system as a model system for studying the interplay between preferences and
norms in \textsc{Mas} architectures. We hope that it will be possible to
transform the definition of a \textsc{Dalmas} into a system that allows for
study through microsimulation (cf.~\cite{BoHo03}).

\begin{definitiontest}
A \textsc{Dalmas} is an ordered 7-tuple $\langle \Omega ,S,A,\mathcal{A}
,\Delta ,\Pi ,\Gamma \rangle $ containing

\begin{itemize}
\item an agent set $\Omega $ ($\omega ,\varkappa ,\omega _{1},...$ elements
in $\Omega $),

\item a state or phase space $S$ ($r,s,s_{1},...$ elements in $S$),

\item an action set $A$ such that for all $a\in A$, $a:\Omega \times
S\longrightarrow S$ such that $a(\omega ,r)=s$ means that if the agent $
\omega $ performs the act $a$ in state $r$, then the result will be state $s$
($a,b,a_{1},...$ elements in $A$),\footnote{%
According to Savage~\cite{Sa72} p.13: \textquotedblleft an act is a
function attaching a consequence to each state of the
world\textquotedblright .}

\item a function $\mathcal{A}:\Omega \times S\longrightarrow \wp (A)$ where $%
\wp (A)$ is the power set of $A;$ $\mathcal{A}(\omega ,s)$ is the set of
acts accessible (feasible) for agent $\omega $\ in state $s$,

\item a deontic structure-operator $\Delta :\Omega \times S\longrightarrow 
\mathcal{D}$ where $\mathcal{D}$\ is a set of deontic structures of the same
type with subsets of $A$ as domains and $\Delta (\omega ,s)$ is $\omega $'s
deontic structure on $\mathcal{A}(\omega ,s)$ in state $s$,\footnote{%
Two structures are of the same type if they have the same number of
relations and corresponding relations in both structures have the same arity.%
}

\item a preference structure-operator $\Pi :\Omega \times S\longrightarrow 
\mathcal{P}$ where $\mathcal{P}$\ is a set of preference structures of the
same type with subsets of $A$ as domains and $\Pi (\omega ,s)$ is $\omega $
's preference structure on $\mathcal{A}(\omega ,s)$ in state $s$,

\item a choice-set function $\Gamma :\Omega \times S\longrightarrow \wp (A)$
where$\mathcal{\ }\Gamma (\omega ,s)$ is the set of actions for $\omega $ to
choose from in state $s$.
\end{itemize}
\end{definitiontest}\bigskip

A \emph{situation} for the \textsc{Dalmas} $\mathfrak{D}$ is determined by the agent to
move, $\omega $, and the state $s$. A situation is represented by an ordered
pair $\langle \omega ,s\rangle .$ The set of situations for $\mathfrak{D}$
is thus $\Omega \times S.$

In a \textsc{Dalmas}, all the agents have the same initial set of actions.
The set of actions to choose from (the choice-set) in a situation $
\left\langle \omega ,s\right\rangle $ is determined by the agent's deontic
structure $\Delta (\omega ,s)$ and preference structure $\Pi (\omega ,s)$.
If $\Gamma
(\omega ,s)$\ consists of one action, then this action applied in the
situation $\left\langle \omega ,s\right\rangle $ is the\ resulting state when $\omega $\
acts in state $s$.

A \emph{simple} \textsc{Dalmas} is a \textsc{Dalmas} containing the
following simple versions of $\Delta $, $\Pi $, and $\Gamma .$

\begin{enumerate}
\item $\Delta (\omega ,s)\subseteq \mathcal{A}(\omega ,s)$ and $\Delta 
\mathcal{(}\omega ,s)$ is the set of permissible actions for $\omega $ in
the state $s$,

\item $\Pi (\omega ,s)=\langle \mathcal{A}(\omega ,s),\succsim \rangle $
where $\succsim $\ is a weak ordering,\footnote{$\left\langle A,\succsim
\right\rangle $ is a \emph{weak ordering} if for all $a,b,c$ in $A$, the
following axioms are satisfied: $a\succsim b$ or $b\succsim a$; 
if $a\succsim b$ and $b\succsim c$, then $a\succsim c$.}

\item $\Gamma (\omega ,s)=\left\{ x\in \Delta (\omega ,s):\text{ for all }
y\in \Delta (\omega ,s),\text{ }x\succsim y\right\} .$
\end{enumerate}

Hence, in a simple \textsc{Dalmas} the choice-set consists of the best
actions which are permissible. Among the elements in $A$ there can be a pass
action, which means the agent does nothing. If we combine the existence of
such an action with very short clock cycles, we obtain systems with close to
asynchronous behaviour (cf.~\cite{BoHo03}~\cite{HuGl93}). 

A \textsc{Dalmas} is not
deterministic, since it does not determine in which order the agents are
going to move, and the choice-set may contain more than one action in every
situation. Let us therefore make the following definition.\smallskip

\begin{definitiontest}
A \emph{deterministic} \textsc{Dalmas} is an ordered 9-tuple

$\langle \Omega,A,S,\mathcal{A},\Delta ,\Pi ,\Gamma ,\tau ,\gamma \rangle $ such that $
\langle \Omega ,A,S,\mathcal{A},\Delta ,\Pi ,\Gamma \rangle $ is a \textsc{
Dalmas} and 

\begin{itemize}
\item $\tau :\Omega \longrightarrow \Omega $ is a turn-operator such that $%
\tau (\omega )=\varkappa $ means that it is $\varkappa $'s turn after $%
\omega $; $\tau $ determines a simple agent priority,

\item $\gamma :\wp (A)\longrightarrow A$ is a tie-breaking function,
determining which of several permissible and equally preferred actions to
choose.
\end{itemize}
\end{definitiontest}
\smallskip 

Note that $\gamma (\Gamma (\omega ,s))\in A$ and thus $\left[ \gamma (\Gamma
(\omega ,s))\right] (\omega ,s)\in S.$ Define $f:\Omega \times
S\longrightarrow \Omega \times S$ in the following way:\smallskip 

$f(\omega
,s)=\langle \tau (\omega ),\left[ \gamma (\Gamma (\omega ,s)) \right]
(\omega ,s)\rangle .$
\smallskip 
Note further that it is possible to iterate $f.$ We define $f^{n}$ as:

$f^{1}(\omega ,s)=f(\omega ,s)$

$f^{n}(\omega ,s)=f(f^{(n-1)}(\omega ,s)).$ \smallskip

\begin{definitiontest}
The $k$-event \emph{run} of a deterministic \textsc{Dalmas} $\mathfrak{D}$ determined
by the \emph{initial situation} $\left\langle \omega _{0},s_{0}\right\rangle 
$ is the sequence $\left\langle \left\langle \omega _{0},s_{0}\right\rangle
,f^{1}(\omega _{0},s_{0}),...,f^{k}(\omega _{0},s_{0})\right\rangle $.
\end{definitiontest}

Suppose that $\langle \Omega ,A,S,\mathcal{A},\Delta ,\Pi ,\Gamma ,\tau
,\gamma \rangle $ is a deterministic \textsc{Dalmas}. Then $\langle \Omega
\times S,T,\phi \rangle $ is a dynamical system, where $T$ is the set of
natural numbers and $\phi $ is the evolution operator defined by $\phi
:\Omega \times S\times T\longrightarrow \Omega \times S$ such that\smallskip

$\phi (\omega ,s,0)=\left\langle \omega ,s\right\rangle $

$\phi (\omega ,s,t)=f^{t}(\omega ,s)$ if $t\geq 1.$\footnote{%
See, for example, \cite{Br91} p.8 for a definition of dynamical system.}\smallskip 
\newline
Note that $\left\langle \phi (\omega _{0},s_{0},0),\phi (\omega
_{0},s_{0},1),...,\phi (\omega _{0},s_{0},k)\right\rangle $ is the run of $%
\mathfrak{D}$ determined by $\left\langle \omega _{0},s_{0}\right\rangle $
and consisting of $k$ moves.

\subsection{The Waste Collector System as a \textsc{Dalmas}}

We elucidate some of the aspects of the definition of a \textsc{Dalmas}
using the waste collectors example introduced in section 2.1.

\subsubsection{Agents and States}

Suppose that we have a set $\Omega $ of $k$ agents, $\Omega =\left\{ \omega
_{1},...,\omega _{k}\right\} .$ A state for the system is characterized by
the position of each of the agents, the location of the waste, and the
amount of waste each agent has collected. A state $s $ is therefore
characterized by three functions,

$\pi :\Omega \longrightarrow $N$^{2}$ such that $\pi (\omega _{1})\neq \pi
(\omega _{2})$ if $\omega _{1}\neq \omega _{2}.$

$\gamma :$ N$^{2}\longrightarrow$ Re.

$\sigma :\Omega \longrightarrow$ Re.

$\pi $ assigns a position to each of the agents, $\gamma $ assigns the
amount of waste to each position, and $\sigma $ assigns the amount of waste
each agent has collected. To denote points in N$^{2}$, bold face letters $%
\mathbf{x}$,$\mathbf{y}$,... will be used. If $\mathbf{x=} \left\langle
x_{1},x_{2}\right\rangle \in$ N$^{2}$ then $\mathbf{x}^{(1)}=x_{1}$
and $\mathbf{x}^{(2)}=x_{2}.$ If the state $s$ is represented by the ordered
triple of $\pi $,$\gamma $ and $\sigma $, i.e. $s=\left\langle \pi ,\gamma
,\sigma \right\rangle ,$ then $s^{(1)}=\pi $, $s^{(2)}=\gamma $, and $%
s^{(3)}=\sigma .$ Let $S$ be the state space of the system. A state $s$ can be an initial
state if $s^{(3)}(\omega )=0$, for all $\omega \in \Omega .$

\subsubsection{Actions and Utility}

Each agent has a repertoire of four different actions, viz. \emph{going
east, going south, going west}, and \emph{going north}. For simplicity, we
denote the actions with their point of compass: \emph{east, south, west},
and \emph{north.} Now, \emph{east} is defined as: \smallskip $east(\omega
,s)=\left\langle \pi ,\gamma ,\sigma \right\rangle $ where

\begin{enumerate}
\item $\pi (\omega )=\langle (s^{(1)}(\omega ))^{(1)}+1,(s^{(2)}(\omega
))^{(2)}\rangle $ and $\pi (\varkappa )=s^{(1)}(\varkappa )$, if $\varkappa
\neq \omega ,$

\item $\gamma (x,y)=0$ if $\pi (\omega )=\left\langle x,y\right\rangle $ and 
$\gamma (x,y)=s^{(2)}(x,y)$, if $\pi (\omega )\neq \left\langle
x,y\right\rangle ,$

\item $\sigma (\omega )=s^{(3)}(\omega )+s^{(2)}(\pi (\omega )).$
\end{enumerate}

\noindent The other actions are defined analogously. The `feasibility
function'\ $\mathcal{A}$ is such that an action is feasible as long
as the performing of the action leads to a new state within the boundaries
of the grid.

The preference structure is determined by a utility function 

$U_{\omega
}:S\longrightarrow$ Re for every agent $\omega $: $\Pi (\omega
,s)=\left\langle \mathcal{A}(\omega ,s),\succsim _{\omega ,s}\right\rangle $%
, such that $a\succsim _{\omega ,s}b$ iff $U_{\omega }(a(\omega ,s))\geq
U_{\omega }(b(\omega ,s)).$

The utility function for an agent can be defined in many ways and we will
leave open the exact definition of it. The norm-regulated \textsc{Dalmas} 
can have its choice-set function $\Gamma$ defined as: 

$\Gamma 
(\omega ,s)=\left\{ x\in \Delta (\omega ,s):\text{ for all }y\in 
\Delta(\omega ,s),\text{ }x\succsim _{\omega ,s}y\right\} .$

\section{Conditions and Normative Systems for a \textsc{Dalmas}}

\subsection{Preamble}

In this section, the building blocks of norms which can regulate a \textsc{%
Dalmas} will be described. The grounds and consequences consist of conditions
on agents, true or false in a situation. This kind of conditions are called
sit-conditions. As its core, a sit-condition has a state-condition, i.e. a
condition true or false in a state. We obtain sit-conditions from
state-conditions by the application of the operators $M,T_{1},...,T_{7}.$
Suppose $c$ is a $\nu $-ary state-condition. Then $c(\omega _{1},...,\omega
_{k})$ is true or false in a state $s$ and $Mc$ and $T_{i}c$ are $\nu $%
+1-ary sit-conditions. $Mc(\omega _{1},...,\omega _{\nu },\omega _{\nu +1})$
is defined as being true in situation $\left\langle \omega ,s\right\rangle $
if and only if $c(\omega _{1},...,\omega _{k})$ is true in $s$ and it is $%
\omega _{\nu +1}$'s turn to draw, i.e. $\omega _{\nu +1}=\omega $. $%
T_{i}c(\omega _{1},...,\omega _{\nu },\omega _{\nu +1})$ is true in $%
\left\langle \omega ,s\right\rangle $ if and only if \textbf{T}$_{i}$ is
the normative position for $\omega _{\nu +1}$ with respect to
$c(\omega _{1},...,\omega _{\nu })$ being true in the state that will be
the result of the action taken in $\left\langle \omega ,s\right\rangle $, 
where \textbf{T}$_{i}$\ is the $i$th type of the
one agent positions. Elementary norms for a \textsc{Dalmas} are ordered pairs of the form $%
\left\langle Mc,T_{i}d\right\rangle $ where $c$ and $d$ are
state-conditions, $Mc$\ is the ground, and $T_{i}d$\ the consequence.
In this essay, we will focus on elementary norms.

The rest of this section is devoted to a detailed development of the ideas
just presented.

\subsection{State-Conditions and Sit-Conditions}

The idea behind the definition of a \textsc{Dalmas} is that its behaviour
will be regulated by a normative system and that the normative system will
be represented by a ground-consequence system $\left\langle \mathcal{B}_{1},%
\mathcal{B}_{2},\mathcal{J}\right\rangle $ where $\mathcal{B}_{1}$ and $%
\mathcal{B}_{2}$\ are condition implication structures. The condition of
interests in connection with a \textsc{Dalmas} is a little more complicated
than those discussed in section 2\ and it is pertinent to distinguish
between two different kinds of conditions, viz.

\begin{enumerate}
\item condition on agents in a state, abbreviated just `state-condition'

\item condition on agents in a situation, abbreviated just `sit-condition'.
\end{enumerate}

A $\nu -$ary state-condition $c$ is true or false of $\nu $ agents $\omega
_{1},...,\omega _{\nu }$ in a state $s$, which will be written $c(\omega
_{1},...,\omega _{\nu };s).$ Note the use of the semicolon to separate the
state from the agents. A $\nu -$ary sit-condition $d$ is true or false of $%
\nu $ agents $\omega _{1},...,\omega _{\nu }$ in a state $\left\langle
\omega ,s\right\rangle $, which will be written $d(\omega _{1},...,\omega
_{\nu };\omega ,s).$

A normative system regulating a \textsc{Dalmas} is a ground-consequence
system $\left\langle \mathcal{B}_{1},\mathcal{B}_{2},\mathcal{J}%
\right\rangle $ where $\mathcal{B}_{1}$ and $\mathcal{B}_{2}$\ are condition
implication structures consisting of certain kinds of sit-conditions. These
sit-conditions are the result of applying certain operations to
state-conditions. We discuss this in next subsection. 

\subsection{The Move-Operator}

Let us now introduce the move-operator $M$ which transforms a $\nu -$ary
state-condition to a $\nu -$ary sit-condition.\smallskip

\begin{definitiontest}
$M$ is an operator on state-conditions such that if $c$ is a $\nu -$ary
state-condition then $Mc$ is a $\nu -$ary sit-condition and\newline
$Mc(\omega _{1},...,\omega _{\nu },\omega _{\nu +1};\omega ,s)$ iff $\omega
_{\nu +1}=\omega $ and $c(\omega _{1},...,\omega _{\nu };s).$
\end{definitiontest}\bigskip

Note that $Mc(\omega _{1},...,\omega _{\nu },\omega _{\nu +1};\omega ,s)$
means that $\omega _{\nu +1}$ is to move in state $\left\langle \omega
,s\right\rangle $ (since $\omega _{\nu +1}=\omega )$, and $c(\omega
_{1},...,\omega _{\nu };s).$ 

If $\langle B,\wedge ,^{\prime },R\rangle $ is a Boolean quasi-ordering
where $B$ consists of state-conditions. Define $B_{M}=\left\{ Mb:b\in B\right\} .$
We can now define a unary operation $_{M}^{\prime }$ on $B_{M}$ in the following
way: $(Mc)_{M}^{\prime }=M(c^{\prime })$. 

Note that $(Mc)_{M}^{\prime }
(\omega _{1},...,\omega _{\nu },\omega _{\nu +1};\omega ,s)$ 
iff 
$M(c^{\prime })(\omega _{1},...,\omega _{\nu },\omega _{\nu +1};\omega ,s)$
iff $\omega _{\nu +1}=\omega $ and $c^{\prime
}(\omega _{1},...,\omega _{\nu };s).$
Define a binary relation $\wedge _{M}$ on $B_{M}$\ in the following way: 
$ Mb\wedge _{M}Mc=M(b\wedge c).$

Note that $(Mb\wedge_{M} Mc)(\omega _{1},...,\omega _{\nu },\omega _{\nu
+1};\omega ,s)$ iff 

$M(b\wedge c)(\omega _{1},...,\omega _{\nu },\omega _{\nu +1};\omega ,s)$
iff 

$\left[ \omega _{\nu +1}=\omega \text{ and }(b\wedge c)(\omega
_{1},...,\omega _{\nu };s)\right] $
iff 

$\left[ \omega _{\nu +1}=\omega \text{ and }b(\omega _{1},...,\omega
_{\nu };s)\text{ and }c(\omega _{1},...,\omega _{\nu };s)\right] $
iff 

$\left[ (Mb)(\omega _{1},...,\omega _{\nu },\omega _{\nu +1};\omega ,s)
\text{ and }(Mc)(\omega _{1},...,\omega _{\nu },\omega _{\nu +1};\omega ,s)
\right]$.
\newline
Let us further define a binary relation $R_{M}$ on $B$ in the following way: 
$bRc.$ It is easy to see that $M$ is an isomorphism from $\langle
B,\wedge ,^{\prime },R\rangle $ to $\langle B_{M},\wedge_{M},^{\prime
}_{M},R_{M}\rangle $ which thus also is a Boolean quasiordering. We say that $%
\langle B_{M},\wedge_{M},^{\prime }_{M},R_{M}\rangle $ is the \emph{m-cis} over 
$\langle B,\wedge ,^{\prime },R\rangle .$

\subsection{The Type-Operators}

Type-operators (see~\cite{LiOd02}) can be applied to state-conditions,
resulting in sit-conditions.\smallskip

\begin{definitiontest}
For $i$, $1\leq i\leq 7,$ $T_{i}$ is an operator on state-conditions such
that if $c$ is a $\nu -$ary state-condition then $T_{i}c$ is a $\nu -$ary
sit-condition and $T_{i}c(\omega _{1},...,\omega _{\nu },\omega _{\nu
+1};\omega ,s)$ iff \textbf{T}$_{i}(\omega _{\nu +1},c(\omega
_{1},...,\omega _{\nu };s^{+}))$
where \textbf{T}$_{i}$ is the $i$th type of the one agent positions and $%
s^{+}$ is the state which will be the result of the action taken by $\omega $%
\ in state $s$.
\end{definitiontest}\bigskip

The meaning of \textbf{T}$_{i}$ is discussed in more detail in section 5.

Suppose that $\langle B,\wedge ,^{\prime },R\rangle $ is a Boolean
quasi-ordering where $B$ consists of state-conditions, define
$B_{T}=\{T_{i}b:b\in B$ and $1\leq i\leq 7\}.$ \noindent $B_{T}$ 
is the set of \emph{normative positions} over $B$. 
If $T_{i}b,T_{j}c \in B_{T}$, then define

(1) $(T_{i}b \wedge_{T} T_{j}c)(\omega _{1},...,\omega _{\nu },\omega _{\nu
+1};\omega ,s)$ iff $T_{i}b(\omega _{1},...,\omega _{\nu },\omega _{\nu
+1};\omega ,s)$ and $T_{j}c(\omega _{1},...,\omega _{\nu },\omega _{\nu
+1};\omega ,s)$,

(2) $(T_{i}b)^{\prime }_{T}(\omega _{1},...,\omega _{\nu },\omega _{\nu
+1};\omega ,s)$ iff $\neg T_{i}b(\omega _{1},...,\omega _{\nu },\omega _{\nu
+1};\omega ,s)$.
\bigskip

Define $B_{T}^{\ast }$ recursively as follows:\smallskip

(1) $B_{T}\subseteq B_{T}^{\ast }$

(2) If $p,q\in B_{T}^{\ast }$ then $p^{\prime }_{T}\in B_{T}^{\ast }$ and $%
p\wedge_{T} q\in B_{T}^{\ast }$

(3) The only members of $B_{T}^{\ast }$ are those resulting from a finite
number of applications of (1) and (2).\smallskip

\begin{definitiontest}
A \emph{normative-position-cis (np-cis)} over a cis $\langle B,\wedge
,^{\prime },R\rangle $ is a Boolean quasi-ordering $\langle B_{T}^{\ast
},\wedge_{T} ,^{\prime }_{T},R_{T}\rangle $ with ${\large \top }_{T}$ as unit
element, ${\large \bot }_{T}$ as zero element and where $Q_{T}$ is the
indifference part of $R_{T}$ such that the following
requirements are satisfied. For any $c,d\in B$ it holds that:\newline
(1) if $i\neq j,$ then $(T_{i}d\wedge_{T} T_{j}d) R_{T} {\large \bot }_{T}\;($for 
$i,j\in \{1,...,7\}),$\newline
(2) ${\large \top }_{T}R_{T}(T_{1}d\vee_{T} ...\vee_{T} T_{7}d)$, \newline
(3) $T_{1}d Q_{T} T_{1}(d^{\prime }),$ $T_{3}d Q_{T} T_{3}(d^{\prime }),$ $%
T_{6}d Q_{T} T_{6}(d^{\prime }),$ $T_{2}d Q_{T} T_{4}(d^{\prime }),$ $%
T_{5}d Q_{T} T_{7}(d^{\prime })$\newline
(4) if $c=d$ then $T_{i}c Q_{T} T_{i}d\ ($for $i,j\in \{1,...,7\}),$\newline
(5) $T_{i}(\top ) Q_{T} \bot _{T}$ if $i=1,3,4,7,\newline
$(6) $T_{i}(\bot ) Q_{T} \bot _{T}$ if $i=1,2,3,5.$
\end{definitiontest}\bigskip

The reason for (3) is a kind of \textquoteleft symmetry
principle \textquoteright (see section 2.3), and the justification for (4) is a logical
postulate. The notion of an np-cis was introduced in \cite{LiOd02}.

\subsection{A Norm-Regulated \textsc{Dalmas}}

\begin{definitiontest}
A \emph{normative system} $\mathcal{N}$ for a \textsc{Dalmas} $\mathfrak{D}$
is a gc-system $\left\langle \mathcal{C}_{M},\mathcal{D}_{T},\mathcal{J}%
\right\rangle $ such that $\mathcal{C}_{M}$=$\left\langle C_{M},\wedge_{M},
^{\prime }_{M},R_{M}\right\rangle $ is the \emph{m-cis} over $\left\langle C,\wedge
,^{\prime },R_{C}\right\rangle $ and $\mathcal{D}_{T}=\langle D_{T}^{\ast
},\wedge_{T} ,^{\prime }_{T},R_{T}\rangle $ is an np-cis over $\langle D,\wedge
,^{\prime },R_{D}\rangle $\ where $C$\ and $D$\ are sets of state-conditions for 
$\mathfrak{D}$.
\end{definitiontest}\bigskip

In the sequel, we will omit the $M$ and $T$ subscripts in 
$\wedge_{M}$ and $^{\prime }_{M}$, and in $\wedge_{T}$ and $^{\prime }_{T}$.

If $\left\langle \mathcal{C}_{M},\mathcal{D}_{T},\mathcal{J}\right\rangle $\
is a normative system, then it is joining $\mathcal{C}_{M}$\ and $\mathcal{D}%
_{T}$. If $\left\langle x,y\right\rangle \in \mathcal{J}$ then $\left\langle
x,y\right\rangle $\ is a \emph{norm} in $\mathcal{N}$ and $x$ is \emph{the
ground} and $y$ is \emph{the consequence} of that norm. Note that a norm in $%
\mathcal{N}$ is a correlation of normative consequences to descriptive
conditions. If $\left\langle x,y\right\rangle $ is a norm in $\mathcal{N}$
and $y\in D_{T}$ then $\left\langle x,y\right\rangle $ is an \emph{%
elementary norm}. An elementary norm in $\mathcal{N}$ is an ordered pair $%
\left\langle Mc,T_{i}d\right\rangle $ where $c\in C$ and $d\in D$ and the
intended interpretation of it\ is a sentence of the following
form:\smallskip \newline
$\forall \omega _{1},...,\omega _{\nu },\omega \in \Omega :\forall s\in
S:Mc(\omega _{1},...,\omega _{\nu },\omega ;\omega ,s)\rightarrow
T_{i}d(\omega _{1},...,\omega _{\nu },\omega ;\omega ,s^{+})$\smallskip 
\newline
or somewhat more generally:\newline
\smallskip $\forall \omega _{1},...,\omega _{\varphi },\omega \in \Omega
:\forall s\in S:Mc(\omega _{1},...,\omega _{\mu },\omega ;\omega
,s)\rightarrow T_{i}d(\omega _{1},...,\omega _{\nu },\omega ;\omega ,s^{+})$%
\newline
where $\varphi =\max \{\mu ,\nu \}$ and $s^{+}$ is the state which will be
the result of the action taken by $\omega $\ in state $s$. In both cases the
norm above is represented as the ordered pair $\left\langle
Mc,T_{i}d\right\rangle .$
\newline
\begin{definitiontest}
A \emph{norm-regulated }\textsc{Dalmas} is a system $[\mathfrak{D,}\mathcal{N%
}]$ where $\mathfrak{D}$ is a \textsc{Dalmas} and $\mathcal{N}$ is a
normative system for $\mathfrak{D}.$
\end{definitiontest}\bigskip

The idea behind a norm-regulated \textsc{Dalmas} is that the deontic
structure operator $\Delta $ is defined in terms of $\mathcal{N}$ in the
sense that what is permissible to do in a situation is determined by the
normative system. We will return to this in section 5.

\subsection{Norms for the Waste Collectors}

Essential to the behaviour of the collectors is how close they are to each
other. To be able to talk about the nearness of the collectors it is
convenient to infer the notion of an $n$-surrounding.\smallskip

\begin{definitiontest}
The $n-$\emph{surrounding} of the point $\langle x,y\rangle \in $ N$\,^{2}$
is the set $\{\langle z,u\rangle :|x-z|\leq n$ and $|y-u|\leq n\}$, and denoted
by $Surr_{n}(x,y).$
\end{definitiontest}\bigskip

We can now define a family of state-conditions expressing overlapping
surroundings:\smallskip

\begin{definitiontest}
$Lap_{j}$ is the state-condition such that $Lap_{j}(\omega _{1},\omega
_{2};s)$ iff $Surr_{j}(s^{(1)}(\omega _{1}))\cap Surr_{j}(s^{(1)}(\omega
_{2}))$ contains exactly $j$ elements.
\end{definitiontest}\bigskip

If we apply the move operator and the type operators, we obtain
sit-conditions from the state-conditions $Lap_{j}$, i.e. we get $MLap_{j}$
and $T_{i}Lap_{j}.$We also need the non-identity condition $d,$ defined by $
d(\omega _{1},\omega _{2})$ iff $\omega _{1}\neq \omega _{2},$ and the
sit-condition $Md$.

We can use the introduced terminology to express the norms for the waste
collectors presented in subsection 2.1. The result is the following set of
ordered pairs of sit-conditions. For legibility reasons, we use $-$ instead of
$\prime$ for negation.

\begin{enumerate}
\item $\left\langle MLap_{0},T_{4}Lap_{2}\vee T_{6}Lap_{2}\vee
T_{7}Lap_{2}\right\rangle $

\item $\left\langle MLap_{0},T_{4}Lap_{3}\vee T_{6}Lap_{3}\vee
T_{7}Lap_{3}\right\rangle $

\item $\left\langle MLap_{0},T_{1}Lap_{0}\vee T_{2}Lap_{0}\vee
T_{3}Lap_{0}\vee T_{5}Lap_{0}\right\rangle $

\item $\left\langle MLap_{0},T_{1}Lap_{1}\vee T_{2}Lap_{1}\vee
T_{3}Lap_{1}\vee T_{5}Lap_{1}\right\rangle $

\item For all $i,0\leq j\leq 9,$ $\left\langle MLap_{1},T_{1}Lap_{j}\vee
T_{2}Lap_{j}\vee T_{3}Lap_{j}\vee T_{5}Lap_{j}\right\rangle $

\item For all $i,0\leq j\leq 9,$ $\left\langle MLap_{2},T_{1}Lap_{j}\vee
T_{2}Lap_{j}\vee T_{3}Lap_{j}\vee T_{5}Lap_{j}\right\rangle $

\item $\left\langle M(-Lap_{4}\wedge -Lap_{6}\wedge
-Lap_{9}),T_{7}Lap_{6}\right\rangle $

\item $\left\langle MLap_{4},T_{7}Lap_{3}\right\rangle $

\item $\left\langle MLap_{6},T_{5}(Lap_{4}\vee Lap_{6}\vee
Lap_{9})\right\rangle $

\item $\left\langle Md,T_{7}Lap_{9}\right\rangle $

\item $\left\langle M\top ,T_{1}Lap_{0}\vee T_{2}Lap_{0}\vee
T_{3}Lap_{0}\vee T_{5}Lap_{0}\right\rangle $
\end{enumerate}

The correspondence between the norms $(n_{1})-(n_{8})$ in section 2 and the
norms above is the following:\smallskip

$(n_{1})$ is represented as (1) and (2),

$(n_{2})$ is represented as (3) and (4),

$(n_{3})$ is represented as (5) and (6),

$(n_{4})$ is represented as (7),

$(n_{5})$ is represented as (8),

$(n_{6})$ is represented as (9),

$(n_{7})$ is represented as (10),

$(n_{8})$ is represented as (11).\smallskip

Note that the among the norms (1)-(11), only (7)-(10) are elementary.

\section{From Norms to Actions}

In this section, a scheme for how normative positions will restrict the set
of acts that the agents are permitted to choose from is presented.
The idea behind a norm-regulated \textsc{Dalmas} is that the deontic
structure operator $\Delta $ is defined in terms of $\mathcal{N}$ in the
sense that what is permissible to do in a situation is determined by the
normative system. More formally, let us suppose that the system is in
situation $\left\langle \omega ,s\right\rangle $ and that $\left\langle
Mc,T_{i}d\right\rangle $ is a norm in $\mathcal{N}$. Suppose further that $c$
and $d$ are $\upsilon -$ary and $c(\omega _{1},...,\omega _{\nu };s).$
Hence, $Mc(\omega _{1},...,\omega _{\nu },\omega ;\omega ,s).$ From $%
Mc(\omega _{1},...,\omega _{\nu },\omega ;\omega ,s)$ together with the norm 
$\left\langle Mc,T_{i}d\right\rangle $ follows that $T_{i}d(\omega
_{1},...,\omega _{\nu },\omega ;\omega ,s)$. An important question now is
what restrictions on the set of feasible acts follow from $T_{i}d(\omega
_{1},...,\omega _{\nu },\omega ;\omega ,s),$ i.e. what is prohibited by $%
T_{i}d(\omega _{1},...,\omega _{\nu },\omega ;\omega ,s)$? Let us see what
holds when $i=2$. The intended interpretation of $T_{2}$ is given by

\begin{center}
$T_{2}d(\omega _{1},...,\omega _{\nu },\omega ;\omega ,s)$ iff \textbf{T}$%
_{2}(\omega ,d(\omega _{1},...,\omega _{\nu };s^{+}))$ iff\smallskip

MayDo$(\omega ,d(\omega _{1},...,\omega _{\nu };s^{+}))\;\&\;$MayPass$%
(\omega ,d(\omega _{1},...,\omega _{\nu };s^{+}))\;\&\smallskip $

$\lnot $MayDo$(\omega ,\lnot d(\omega _{1},...,\omega _{\nu };s^{+})).$
\end{center}

Only the third conjunction above results in a prohibition: in the situation $%
\left\langle \omega ,s\right\rangle,$ $\omega $\ may not see to it that not 
$d(\omega _{1},...,\omega _{\nu })$ will be the case in the next state$.$
What does this mean? If in state $s$ $d(\omega _{1},...,\omega _{\nu })$ is
true but in state $s^{+},$ which is the result of $\omega $'s performing
action $a$ in situation $s$, $d(\omega _{1},...,\omega _{\nu })$ is not
true, then $\omega $ has seen to it that not $d(\omega _{1},...,\omega _{\nu
})$, by doing $a$. Since $\omega $ may not see to it that $d(\omega
_{1},...,\omega _{\nu }),$ it follows that $a$ is prohibited. 

Let us now
consider the case $i=3$. According to the intended interpretation of $T_{3}d$%
\ it holds that

\begin{center}
$T_{3}d(\omega _{1},...,\omega _{\nu },\omega ;\omega ,s)$ iff\smallskip

MayDo$(\omega ,d(\omega _{1},...,\omega _{\nu };s^{+}))\;\&\;\lnot $ MayPass$%
(\omega ,d(\omega _{1},...,\omega _{\nu };s^{+}))\;\&\smallskip $

MayDo$(\omega ,\lnot d(\omega _{1},...,\omega _{\nu };s^{+})).$
\end{center}

\noindent If $d(\omega
_{1},...,\omega _{\nu })$ is true in $s$ and $d(\omega _{1},...,\omega _{\nu
})$ is true in $s^{+}$, then Pass$(\omega ,d(\omega _{1},...,\omega _{\nu
};s^{+})).$ If $\lnot d(\omega _{1},...,\omega _{\nu })$ is true in $s$ and $%
\lnot d(\omega _{1},...,\omega _{\nu })$ is true in $s^{+}$\, then Pass$(\omega
,d(\omega _{1},...,\omega _{\nu };s^{+})).$ Therefore, 

Pass$(\omega
,d(\omega _{1},...,\omega _{\nu };s))$ iff $\left[ d(\omega _{1},...,\omega
_{\nu })\text{ is true in }s\text{ iff }d(\omega _{1},...,\omega _{\nu })%
\text{ is true in }s^{+}\right] .$ Hence, if $\omega $ may not be passive
with regard to $d(\omega _{1},...,\omega _{\nu })$ in $\left\langle \omega
,s\right\rangle $, then $[d(\omega _{1},...,\omega _{\nu };s)$ iff $d(\omega
_{1},...,\omega _{\nu };a(\omega ,s))]$ implies that $a$ is prohibited. 

The
other type-operators $T_{i}$ can be analyzed in an analogous way. As the
result of such an analysis, we suggest the following stipulations.\footnote{%
Note that this explication of $T_{1}-T_{7}$ can easily be transformed to
an explication of the one agent types \textbf{T}$_{1}$-\textbf{T}$_{7}$
within the framework of predicate logic.}

\begin{enumerate}
\item From $T_{1}d(\omega _{1},...,\omega _{\nu },\omega ;\omega ,s)$
follows no restriction on the acts.

\item From $T_{2}d(\omega _{1},...,\omega _{\nu },\omega ;\omega ,s)$
follows that\newline
if $d(\omega _{1},...,\omega _{\nu };s)$ and $\lnot d(\omega _{1},...,\omega
_{\nu };a(\omega ,s))$ then Prohibited$_{\omega ,s}(a).$

\item From $T_{3}d(\omega _{1},...,\omega _{\nu }\omega ;\omega ,s)$ follows
that\newline
if $[d(\omega _{1},...,\omega _{\nu };s)$ iff $d(\omega _{1},...,\omega
_{\nu };a(\omega ,s))]$ then Prohibited$_{\omega ,s}(a).$

\item From $T_{4}d(\omega _{1},...,\omega _{\nu },\omega ;\omega ,s)$
follows that\newline
if $\lnot d(\omega _{1},...,\omega _{\nu };s)$ and $d(\omega _{1},...,\omega
_{\nu };a(\omega ,s))$ then Prohibited$_{\omega ,s}(a).$

\item From $T_{5}d(\omega _{1},...,\omega _{\nu },\omega ;\omega ,s)$
follows that\newline
if $\lnot d(\omega _{1},...,\omega _{\nu };a(\omega ,s))$ then Prohibited$
_{\omega ,s}(a).$

\item From $T_{6}d(\omega _{1},...,\omega _{\nu },\omega ;\omega ,s)$
follows that\newline
if not $[d(\omega _{1},...,\omega _{\nu };s)$ iff $d(\omega _{1},...,\omega
_{\nu };a(\omega ,s)]$ then Prohibited$_{\omega ,s}(a).$

\item From $T_{7}d(\omega _{1},...,\omega _{\nu },\omega ;\omega ,s)$
follows that\newline
if $d(\omega _{1},...,\omega _{\nu };a(\omega ,s))$ then Prohibited$_{\omega
,s}(a).$\ 
\end{enumerate}

These stipulations can now be used to define the deontic structure-operator $%
\Delta .$ One possibility is to let $\Delta (\omega ,s)$ be the set of
feasible acts $a$ that are not eliminated as Prohibited$_{\omega ,s}(a)$
according to the rules 1-7 above, where Prohibited$_{\omega ,s}(a)$ is
equivalent to $\lnot $Permissible$_{\omega ,s}(a)$.\footnote{%
The mentioned connection between Prohibited and Permissible is naturally not
the only possible. } Hence, $\Delta
(\omega ,s)=\{$Permissible$_{\omega ,s}(a):a\in A\}.$\medskip

Note that in the outset all feasible acts are permissible, i.e. for all $
a\in A$, Permissible$_{\omega ,s}(a)$. The basic idea is now that we
eliminate elements from the set of permissible\ acts for $\omega $ in $s$
using the norms and sentences expressing what holds for the agents with
respect to grounds in the norms. To facilitate the presentation it is
convenient to introduce the following six operators on state-conditions:

\begin{description}
\item $E_{2}^{a}d(\omega _{1},...,\omega _{\nu },\omega ;\omega ,s)$ iff $%
[d(\omega _{1},...,\omega _{\nu };s)$ and $\lnot d(\omega _{1},...,\omega
_{\nu };a(\omega ,s))]$

\item $E_{3}^{a}d(\omega _{1},...,\omega _{\nu },\omega ;\omega ,s)$ iff $%
[d(\omega _{1},...,\omega _{\nu };s)$ iff $d(\omega _{1},...,\omega _{\nu
};a(\omega ,s))]$

\item $E_{4}^{a}d(\omega _{1},...,\omega _{\nu },\omega ;\omega ,s)$ iff $%
[\lnot d(\omega _{1},...,\omega _{\nu };s)$ and $d(\omega _{1},...,\omega
_{\nu };a(\omega ,s))]$

\item $E_{5}^{a}d(\omega _{1},...,\omega _{\nu },\omega ;\omega ,s)$ iff $%
[\lnot d(\omega _{1},...,\omega _{\nu };a(\omega ,s))]$

\item $E_{6}^{a}d(\omega _{1},...,\omega _{\nu },\omega ;\omega ,s)$ iff not 
$[d(\omega _{1},...,\omega _{\nu };s)$ iff $d(\omega _{1},...,\omega _{\nu
};a(\omega ,s))]$

\item $E_{7}^{a}d(\omega _{1},...,\omega _{\nu },\omega ;\omega ,s)$ iff $%
d(\omega _{1},...,\omega _{\nu };a(\omega ,s)).$
\end{description}

\noindent Note that for all $i,$ $2\leq i\leq 7,$ $(T_{i}d\wedge
E_{i}^{a}d)(\omega _{1},...,\omega _{\nu },\omega ;\omega ,s)$ implies that
Prohibited$_{\omega ,s}(a)$.

In situation $\left\langle \omega ,s\right\rangle $, the action $a$ is
prohibited if there are $c\in C$ and $d\in D$ such that $%
\left\langle Mc,T_{i}d\right\rangle $ is a norm for some $i,$ $2\leq i\leq
7, $ and there are $\omega _{1},\omega _{2},...\omega _{\nu }\in \Omega $
such that $Mc(\omega _{1},...,\omega _{\nu },\omega ;\omega ,s)$ and $%
E_{i}^{a}d(\omega _{1},...,\omega _{\nu },\omega ;\omega ,s).$ The essential
part of the procedure is the following inference:

\begin{center}
\underline{$Mc(\omega _{1},...,\omega _{\nu },\omega ;\omega
,s)\;\&\;\langle Mc,T_{i}d\rangle $} \ \ \ \ \ \ \ \ \ \ \ \ \ \ \ \ \ \ \ \
\ \ \ \ \ \ \ \ \ \ \ \ \ \ \ \ \ \ \ \ \ \ \ \ \ \ \ \ \ 

\underline{ $\ \ \ \ \ \ \ \ T_{i}d(\omega _{1},...,\omega _{\nu
},\omega ;\omega ,s)$ $\ \ \ \ \ \ \ \ \ \ \ \ \ \ \ \ \ \ \ \ \ \ \ \ \ \ \
\ \ \ E_{i}^{a}d(\omega _{1},...,\omega _{\nu },\omega ;\omega ,s)$}

$Prohibited_{\omega ,s}(a)$
\end{center}

\noindent Now we can for the normative system $\mathcal{N}$=$\left\langle 
\mathcal{C}_{M},\mathcal{D}_{T},\mathcal{J}\right\rangle $ define Prohibited$
_{\omega ,s}$ more explicitly as:\medskip

Prohibited$_{\omega ,s}=\{a\in \mathcal{A}(\omega ,s)$ $\ |$ $\ \exists c\in
C:\exists d\in D:\exists i,2\leq i\leq 7:$

\begin{center}
$\left\langle Mc,T_{i}d\right\rangle \in \mathcal{J}\ \&\ \exists \omega
_{1},...\omega _{\nu }\in \Omega :Mc(\omega _{1},...,\omega _{\nu },\omega
;\omega ,s)$ $\&$ $E_{i}^{a}d(\omega _{1},...,\omega _{\nu },\omega ;\omega
,s)\}.$
\end{center}

\noindent Thus, $\Delta (\omega ,s)$ can be defined as $\Delta (\omega ,s)=%
\mathcal{A}(\omega ,s)\backslash $Prohibited$_{\omega ,s}$, i.e. $\Delta
(\omega ,s)=\{a\in \mathcal{A}(\omega ,s)$ $|\ a \notin $Prohibited$_{\omega
,s}\}.$

For a simple \textsc{Dalmas} it is a small step from $\Delta (\omega ,s)$\
to $\Gamma (\omega ,s)$. If $\mathfrak{D}$ is a simple \textsc{Dalmas}, then 
$\Gamma (\omega ,s)=\Pi (\omega ,s)/\Delta (\omega ,s)$, i.e. $\Gamma
(\omega ,s)=\langle \Delta (\omega ,s),\succsim ^{\prime }\rangle$, where $%
\succsim ^{\prime }$\ is the restriction of $\succsim $ to $\Delta (\omega
,s)$.

In the definition of Prohibited$_{\omega ,s}$\ above, we use only norms that
are elementary. Whether this implies substantial limitations will here be
left as an open problem. Consider for example the norm $\left\langle
Mc,t\right\rangle $ where $t\in D_{T}^{\ast }.$ The problem is under what state
condition $e$ the consequence $t$ implies a prohibition of an act. We do not
deal with this problem, but just point out that
\newline

$(T_{i_{1}}d_{1}\vee T_{i_{2}}d_{2})(\omega _{1},...,\omega _{\nu },\omega
;\omega ,s)\longrightarrow $

\begin{center}
$\left[ (E_{i_{1}}^{a}d_{1}\wedge E_{i_{2}}^{a}d_{2})(\omega _{1},...,\omega
_{\nu },\omega ;\omega ,s)\longrightarrow \text{Prohibited}_{\omega ,s} (a) \right]$.
\end{center}

$(T_{i_{1}}d_{1}\wedge T_{i_{2}}d_{2})(\omega _{1},...,\omega _{\nu },\omega
;\omega ,s)\longrightarrow $

\begin{center}
$\left[ (E_{i_{1}}^{a}d_{1}\vee E_{i_{2}}^{a}d_{2})(\omega _{1},...,\omega
_{\nu },\omega ;\omega ,s)\longrightarrow \text{Prohibited}_{\omega ,s} (a)
\right]$.
\end{center}

Another way of dealing with $\left\langle Mc,t\right\rangle $ is to try and
show that all prohibitions following from it also follow from elementary
(and eventually minimal) norms.

\section{Conclusion and Discussion}

The aim of this study is to present a theory of how norms can be used to
regulate the behaviour of multi-agent systems on the assumption that the role
of norms is to define the \textit{Spielraum} for an agent. The theory can be
summarized as follows. The norms for a \textsc{Mas} are regarded as
belonging to a normative system and such a system is represented
algebraically as a ground-consequence system containing a Boolean
quasi-ordering of grounds and a Boolean quasi-ordering of consequences. The
norms are joinings from grounds to consequences, and the specific normative
content of a normative system is given by the set of minimal norms. The
consequences are expressed using operators on conditions corresponding to
the Kanger-Lindahl types of one-agent positions. An important step in the
theory construction is the specification under what circumstances the
sentence $T_{i}d(\omega _{1},...,\omega _{\nu },\omega ;\omega ,s)$\ implies
that an action $a$\ is prohibited for the agent $\omega $\ in the state $s$
(see section 5). An action $a$ is regarded as a function, which is the usual
way of representing an action in decision theory, and $d$ is a $\nu $-ary
condition on agents, true or false in the situation $s$. An abstract
architecture based on the theory of norm-regulation of behaviour is defined,
and a \textsc{Mas} having this architecture is called a norm-regulated 
\textsc{Dalmas}. The system of waste collectors is an example of a such a
norm-regulated \textsc{Dalmas}.

An important tool in the present study is 
the characterization of abstract architectures by the definitions of
set-theoretical predicates. Among the abstract architectures
defined in this way, the most important one is a norm-regulated \textsc{%
Dalmas}, as exemplified in the previous section. This is just the first step
towards a theory of architectures for \textsc{Mas} that restricts the
behaviour of the system using norms. The theory can be developed by defining
a number of set-theoretical predicates that are specifications of the
predicate \textsc{Dalmas}, and we can obtain a hierarchy of predicates with 
\textsc{Dalmas} as its root.\footnote{%
One set-theoretical predicate $\mathfrak{P}_{2}$ is a specification of
another $\mathfrak{P}_{1}$ if the following holds: if $\mathfrak{S}_{2}$ is
a $\mathfrak{P}_{2}$ there is a sub-sequence $\mathfrak{S}_{1}$ of $%
\mathfrak{S}_{2}$ that is a $\mathfrak{P}_{1}$. The relation `to be a
specification of'\ is a partial order.}

There are many refinements and extensions to be made before it is possible
to state the significance of the theory. For instance, the minimal norms
represent the specific normative content of a normative system. It is
therefore a reasonable question, whether, in the definition of Prohibited$%
_{\omega ,s}$, it is sufficient to regard only minimal norms. Note that the
stipulations expressing under what conditions $T_{i}d$ implies that
the act $a$ is prohibited are independent of the structure over the
consequences. The latter, however, is relevant to determining the minimal
norms. 

Moreover, in the definition of Prohibited$_{\omega ,s}$\ above, we use only
norms that are elementary. Whether this implies substantial limitations
ought to be investigated. Consider the norm $\left\langle Mc,t\right\rangle $
where $t\in D_{T}^{\ast }.$ Under what state condition $e$ does the
consequence $t$ imply a prohibition of an act? Is it possible to prove that
all prohibitions following from it also follow from elementary (and minimal)
norms?

We have only considered norms of a rather special kind. The descriptive
conditions are conditions on agents. Let us call them \textquoteleft
agent norms\textquoteright. An example of an elementary agent norm is $%
\left\langle Mc,T_{i}d\right\rangle $, where $c$ and $d$ are conditions on
agents, true or false in a state. However, it is easy to generalize so that $%
c$ and $d$ can be conditions not just on agents but on a set $U$ containing
agents and objects of other kinds, for example physical objects. It seems to
be straightforward to change our definitions so that this alternative will
be accommodated.

Norms of a radically different character than agent norms are norms
determining which are the agent norms for a \textsc{Mas}. Such norms can for
example give competence to some agents to functions as a kind of
\textquoteleft legislators\textquoteright\ and change the agent norms~%
\cite{Sp03}.

In this essay, we have tried to show how theories of normative positions
and normative systems can be used in constructing an architecture for
norm-regulated \textsc{Mas}. However, we have only used a small portion of
the Kanger-Lindahl theory of normative positions, namely the theory of
one-agent types. According to the intended interpretation of the type
operators in section 4,

\begin{center}
$T_{i}c(\omega _{1},...,\omega _{\nu },\omega _{\nu +1};\omega ,s)$
\end{center}

\noindent means that $\omega _{\nu +1}$\ has the normative positions of type 
$T_{i}$\ with respect to the state of affairs that $c(\omega _{1},...,\omega
_{\nu };s^{+}).$ Here $T_{i}$\ is a one-agent type-operator. A two-agent
type-operator $W_{i}$ applied to $c$ will give a condition $W_{i}c$ such that

\begin{center}
$W_{i}c(\omega _{1},...,\omega _{\nu },\omega _{\nu +1},\omega _{\nu
+2};\omega ,s)$
\end{center}

\noindent means that the party $\omega _{\nu +1}$\ has the normative
positions of type $W_{i}$\ versus the counter party $\omega _{\nu +2}$ with
respect to the state of affairs that $c(\omega _{1},...,\omega _{\nu
};\omega ,s^{+}).$ \footnote{See~\cite{Li77}, chapter 4.} The use of two-agent
type-operators in norm-regulated \textsc{Dalmas} will increase the strength
and flexibility of the \textsc{Dalmas} hierarchy of abstract architectures.

In a norm-regulated \textsc{Dalmas}, the normative system plays the role of
constitutive rules, and it is not possible for the agents to break the
norms. In spite of this, it is reasonable to distinguish between
\textquoteleft may\textquoteright\ and \textquoteleft
can\textquoteright . What an agent $\omega $ \emph{can} do in a situation 
$s$ is given by $\mathcal{A}(\omega ,s)$ and what it \emph{may} do is
determined by $\Delta (\omega ,s)$. Note also that the formal framework for
representing norms can be used for different kinds of norms, even regulative
norms. For example, an agent can choose between obeying a regulative law or
breaking it and thereby giving one or several other agents the right (but
possibly not the duty) to punish it, either immediately or when it is its
turn to move. This can easily be expressed using normative positions.

\textsc{Dalmas} and deterministic \textsc{Dalmas} are intended to function
as wide frames for studying norm-regulations of \textsc{Mas} and thus to be
weak abstract architectures. Are they weak enough or are they excluding
something essential? As has been pointed out earlier, for instance, a 
\textsc{Dalmas} has global states and global dynamics. This fact
notwithstanding, the formal definition of a \textsc{Dalmas} is flexible
enough to incorporate a kind of history dependence. One way of doing this is
to introduce a set $P$ of primary states (phases) such that $s\in S$ \ iff \ 
$s=\left\langle p,\left\langle p_{1},...,p_{\nu }\right\rangle \right\rangle 
$, where $p,p_{1},...,p_{\nu }\in P.$ The intended interpretation is that
when $\mathfrak{D}$ is in $s$ then $\mathfrak{D}$ is in the primary state $p$
and its history is $\left\langle p_{1},...,p_{\nu }\right\rangle $, i.e.
the passed primary states of $\mathfrak{D}$ is given by $\left\langle
p_{1},...,p_{\nu }\right\rangle $. Note that if $s_{1}=\left\langle p,\sigma
_{1}\right\rangle $ and $s_{2}=\left\langle p,\sigma _{2}\right\rangle $
where $\sigma _{1}\neq \sigma _{2},$ then it is possible that $a(\omega
,s_{1})\neq a(\omega ,s_{2}),$ so the result of $\omega $ performing act $a$
in the primary state $p$ is dependent on the passed primary states of $%
\mathfrak{D}$.

Another way to obtain history dependence is to use the definition of a run
(see section 3.1). With some modifications a run can be regarded as a state
consisting of situations based on primary states. In this case we have
dependence not just on passed primary states but on passed primary
situations. If the \textsc{Dalmas} in question is deterministic, passed actions also
determine passed primary situations, but the opposite does not hold.
Therefore, we can distinguish between dependence on previous
primary states, dependence on previous primary situations, and
dependence on previous actions.

Suppose that $\mathfrak{D}$ is a \textsc{Dalmas}, in which the agents
cooperate to solve a problem. Which normative system will lead to the most
effective behavior of the system? It is desirable that $\mathfrak{D}$ itself
could determine the optimal normative system for the task in question. Given
a set of grounds and a set of consequences, which together constitute the
vocabulary of the system, $\mathfrak{D}$ can test all possible sets of
minimal norms. If there is a function for evaluating the result of a run of $%
\mathfrak{D}$,\ then different normative systems can be compared and the
best system can be chosen. A change of vocabulary corresponds to a mutation
in the normative system, which can lead to dramatic changes of efficiency.
Note that, in principle, the evaluation function can be very complicated,
for example it can be multi-dimensional and based on ideas of fairness. This
seems to imply that we can treat multi-agent systems as consisting of
cooperative agents even if they do not have what is usually called
\textquoteleft a common interest\textquoteright\, and cooperate only in the
sense that they want to avoid destructive conflicts. 

\section*{Acknowledgements}
For this research, Jan Odelstad was funded by HSFR (project F1113/97) and the KK-foundation. Magnus Boman gratefully acknowledges Vinnova (Accessible Autonomous Software). The authors would like to express their gratitude to Lars Lindahl for his important input. Finally, the editors and the anonymous referees showed great patience and insight, for which we thank them.

\end{document}